%% file: GFC-main.tex
\documentclass[12pt]{article}
\input definitions.tex

\renewcommand{\theequation}{\arabic{section}.\arabic{equation}}

\begin{document}
\title{
\textsc{Gauge Invariance and Tachyon Condensation}\\
\textsc{in Cubic Superstring Field Theory}
\\
$~$\\}
\author{
\textsf{I.Ya. Aref'eva\footnote{Email: arefeva@mi.ras.ru}},
\,\textsf{D.M. Belov\footnote{Email: belov@orc.ru, belov@mi.ras.ru}},
\,\textsf{A.S. Koshelev\footnote{Email: kas@depni.npi.msu.su}}
\\
\emph{Steklov Mathematical Institute, Russian Academy of Sciences,}\\
\emph{Gubkin st. 8, GSP-1, Moscow, Russia, 117966}
\\
\\
\textsf{and}\\
\textsf{P.B. Medvedev\footnote{Email: medvedev@heron.itep.ru}}\\
\emph{Institute of Theoretical and Experimental Physics,}\\
\emph{B.Cheremushkinskaya st. 25, Moscow, 117218}
}

\date {~}
\maketitle
\thispagestyle{empty}

\begin{abstract}
The gauge invariance of cubic open superstring field theory is
considered in a framework of level truncation, and
applications to the tachyon condensation problem are discussed. As
it is known, in the bosonic case the  Feynman-Siegel
gauge is not universal within the level truncation method.
We explore another gauge
that is more suitable for calculation of the tachyon potential for
fermionic string at level $(2,6)$. We show  that this new gauge has
no restrictions on the region of its validity at least at this
 level.

\end{abstract}

\newpage
\tableofcontents

\section{Introduction}\label{sec:intro}
\setcounter{equation}{0}
\input GFC-intro.tex

\section{General}\label{sec:general}
\setcounter{equation}{0}
\input GFC-general.tex

\section{Calculations of Structure Constants}\label{sec:calc}
\setcounter{equation}{0}
\input GFC-calc.tex

\section{Level Truncation and Gauge Invariance}\label{sec:glts}
\setcounter{equation}{0}
\input GFC-glts.tex

\section{The Orbits}\label{sec:orbits}
\setcounter{equation}{0}
\input GFC-orbits.tex

\section{Summary}\label{sec:summary}
\setcounter{equation}{0}
 We have presented the restricted
cubic superstring action \eqref{action} that contains fields only
from $V_0$ and $V_1$
spaces. One can consider this superselection rule
as a partial gauge fixing.
Then we have described a method
of calculation of $\star$-product without using the
direct expression through Neumann
functions. Using this method we have computed
the gauge transformations of the action truncated
to level $(2,6)$. In contrast to bosonic string field theory
the gauge invariance in level truncation scheme
is broken not only in the second order in coupling constant,
but already in the first order. We have  made an estimation of this
breaking using an example of $t^2$-terms in
gauge variation of the truncated action. It is shown
that this breaking originated
from the contributions of higher level fields is about $33\%$.
This gives us a hope that the gauge invariance
rapidly restores as level grows.

 We have explored a validity
of our gauge condition \eqref{gaugecond} in cubic SSFT as well as
Feynman-Siegel one in boson SFT.
We have shown that
every gauge orbit in cubic SSFT necessarily intersects our
gauge fixing surface \eqref{gaugecond} at least once (see Figure~\ref{pic:super}).
Hence our gauge is valid for computation of extremum of the action.
For Feynman-Siegel gauge in the bosonic SFT we have shown that there are
 field configurations with orbits which never intersect the
gauge fixing surface (see Figure~\ref{pic:boson}).

\section*{Acknowledgments}

This work was supported in part
by RFBR grant 99-61-00166 and by RFBR grant for leading scientific
schools 00-15-96046.
I.A., A.K. and P.M. were supported in part by INTAS grant
99-0590 and D.B. was supported in part by INTAS grant 99-0545.

\newpage
\appendix
\section*{Appendix}
\addcontentsline{toc}{section}{Appendix}
\renewcommand {\theequation}{\thesection.\arabic{equation}}

\section{Notations}
\label{app:notations}
\setcounter{equation}{0}
\input GFCa-notations.tex

\newpage
\section{Conformal Transformations of the Dual Fields}
\label{app:confmap}
\setcounter{equation}{0}
\input GFCa-cmaps.tex


\end{document}

%% file: definitions.tex
\usepackage{amssymb,amsmath,amsthm}
\usepackage{hyperref}
\usepackage{array,longtable}
\usepackage{graphicx}
\usepackage[mathscr]{eucal}

\newcommand{\be}{\begin{equation}}
\newcommand{\ee}{\end{equation}}

\newcolumntype{V}{>{$}m{4cm}<{$}}
\newcolumntype{C}{>{$}c<{$}}
\newcolumntype{L}{>{$}l<{$}}
\newcolumntype{R}{>{$}r<{$}}

\newcommand{\Zh}{\mathbb Z}

\newcommand{\ra}{\rightarrow}

\newcommand{\Ac}{\mathcal{A}}

\newcommand{\pd}{\partial}

\newcommand{\la}{\langle\!\langle}
\renewcommand{\ra}{\rangle\!\rangle}

\newcommand{\df}{{f^{\prime}}}
\newcommand{\ddf}{{f^{\prime\prime}}}
\newcommand{\dddf}{{f^{\prime\prime\prime}}}

\allowdisplaybreaks[3]

%% file: GFC-intro.tex

It has been conjectured by Sen \cite{sen:conjecture} that at the stationary
point of the
tachyon potential for the non-BPS $\mathrm{D}$-brane or
brane-anti-$\mathrm{D}$-brane pair of superstring theories, the
negative energy density cancels the brane tension. Sen's conjecture has been
intensively studied within bosonic SFT \cite{bosonSFT}, non-polynomial NS
SFT \cite{0002211}
as well as in the framework of the boundary conformal field theory
\cite{BCFT}.
We have studied \cite{ABKM} this
conjecture using a cubic superstring field theory
\cite{AMZ1,PTY,AMZ2,Thorn}
extended to GSO$-$ sector. We have computed
the tachyon potential  using the level truncation scheme \cite{KS} at
levels $(1/2,1)$ and $(2,6)$. It is interesting to note that already at
level $(1/2,1)$ one gets $97.5\%$ of the expected result. For
calculations at level $(2,6)$ we have used implicitly a special
gauge.

As it was noted by Ellwood and Taylor \cite{Taylor} there are
restrictions on validity of Feynman-Siegel gauge within the level
truncation method. They have found that this gauge choice breaks
down outside a fairly small region in field space. Moreover, they
have shown that  singularities previously found in the tachyon
effective potential are gauge artifacts arising from the boundary
of the region of validity of Feynman-Siegel gauge.

Therefore a natural question arises whether  the result obtained in
\cite{ABKM} depends on the gauge choice. More technically,
whether the obtained minimum of the tachyon potential  is in a
domain of validity of the chosen gauge. To investigate a
possibility of determination of the local stable vacuum by choosing the
gauge used in \cite{ABKM} we study the orbits of gauge
transformations in the level truncation scheme. On level $2$ the
gauge transformations form a one parametric group and to find
its orbits one has to solve a linear system of first order
differential equations. The  right hand sides of these
differential equations are defined by structure constants of the
level-truncated gauge transformations. To find these structure
constants we use the CFT methods \cite{gluing} which allow us
to calculate the Witten vertex \cite{Witten,GrJe} for given
fields.

Our calculations show that the gauge orbits are arranged in such
a way that our gauge fixing condition intersects all orbits at
least once. This is not the case of
the Feynman-Siegel  in the bosonic theory since there are
orbits which do not intersect the surface specified by
gauge fixing condition.

The paper is organized as follows. Section~\ref{sec:general}
contains a brief review of  extension of the cubic SSFT
to the GSO$-$ sector and a description of specific features
of the level truncation scheme. In Section~\ref{sec:calc} we
perform calculations of the structure constants at level $2$. In
Section~\ref{sec:glts} we discuss the  gauge invariance at  low
levels. In Section~\ref{sec:orbits} we solve equations which
define the orbits of the gauge transformation for boson SFT as
well as for fermionic SFT at level $(2,6)$ and analyze if the  orbits
intersect the gauge fixing surfaces. Appendix~\ref{app:notations}
contains necessary information about our notations and in
Appendix~\ref{app:confmap} we collect conformal transformations of
the dual fields.


%% file: GFC-general.tex
\subsection{Action}
The NS part of the  string field theory on a single
non-BPS D$p$-brane consists of two sectors: GSO$+$ and GSO$-$. Let
us denote by $\Ac_{\pm}$ the string field in GSO$\pm$ sectors.
The string fields $\Ac_{\pm}$ are formal series
\begin{equation}
\Ac_+ = \sum_{i}\int \frac{d^{p+1}k}{(2\pi)^{p+1}}\;\phi^i(k)\Phi_{i}(k),
\qquad
\Ac_-= \sum_{a}\int \frac{d^{p+1}k}{(2\pi)^{p+1}}\;t^a(k)\mathrm{T}_{a}(k).
\end{equation}
which encode infinite number of space-time fields
$\phi^i(k)$ and $t^a(k)$ in momentum representation.
These space-time fields are associated
with zero picture ghost number one
conformal operators $\Phi_i(k)$ and
$\mathrm{T}_a(k)$ of weights $h_i$ and $h_a$.

The cubic NS string field theory action is \cite{ABKM}:
\begin{equation}
\begin{split}
S[\mathcal{A}_+,\mathcal{A}_-]&=-\left[
\frac{1}{2}\la Y_{-2}|\mathcal{A}_+,Q_B\mathcal{A}_+
\rangle\!\rangle+\frac{g_o}{3}\la
Y_{-2}|\mathcal{A}_+,\mathcal{A}_+,\mathcal{A}_+\rangle\!\rangle
\right.\\
&~~~~~~~~~\left.+\frac{1}{2}\la
Y_{-2}|\mathcal{A}_-,Q_B\mathcal{A}_-\rangle\!\rangle -
g_o\la
Y_{-2}|\mathcal{A}_+,\mathcal{A}_-,\mathcal{A}_-\rangle\!\rangle\right].
\end{split}
\label{action}
\end{equation}
Here $Y_{-2}$ is the non-chiral double step inverse
picture changing operator, $g_o$ is a dimensionless
open string coupling constant\footnote{We drop $g_o$
in the majority of formulae}.
$Q_B$ is BRST operator (see Table~\ref{table:2})
and $\la Y_{-2}|\dots\ra$ is the odd bracket:
\begin{align}
\la Y_{-2}| A_1,\dots,A_n\ra=
\left\langle P_n\circ Y_{-2}(0,0)\,F_1^{(n)}\circ A_1(0)\dots
F_n^{(n)}\circ A_n(0)\right\rangle
, \quad n=2,3;
\label{oddbracket}
\\
\intertext{where the maps $P_n$ and $F_j^{(n)}$ are defined as
(see also \eqref{mapsa})}
F^{(n)}_j=P_n\circ f^{(n)}_j,\quad f^{(n)}_j(w)=
e^{\frac{2\pi i}{n}(2-j)}\left(\frac{1+iw}{1-iw}\right)^{2/n},
\quad j=1,\dots,n,\quad n=2,3;
\label{maps}
\\
P_2(z)=i\frac{1-z}{1+z},\quad
P_3(z)=\frac{i}{\sqrt{3}}\frac{1-z}{1+z}.
\end{align}

 The action \eqref{action} is invariant under the
gauge transformations \cite{ABKM}
\begin{equation}
\begin{split}
\delta \mathcal{A}_+&=Q_B\Lambda_++g_o[\mathcal{A}_+,\Lambda_+]
+g_o\{\mathcal{A}_-,\Lambda_-\},\\ \delta
\mathcal{A}_-&=Q_B\Lambda_-+g_o[\mathcal{A}_-,\Lambda_+]
+g_o\{\mathcal{A}_+,\Lambda_-\},
\end{split}
\label{gauge}
\end{equation}
where $[\,,]$ ($\{\,,\}$) denotes $\star$-commutator
(-anticommutator). The Grassman properties of
the fields $\Ac_{\pm}$ and $\Lambda_{\pm}$
are collected in Table~\ref{tab:Grassman}.

\begin{table}[!h]
\begin{center}
\renewcommand{\arraystretch}{1.4}
\begin{tabular}[h]{||C|c|C|C|C||}
\hline \textrm{Name}& Parity & \textrm{GSO} & \textrm{Superghost number}
 & \textrm{Weight}\, (h)\\ \hline
\hline \mathcal{A}_+ & odd & + & 1 & h\in\Zh,\,h\geqslant -1\\
\hline \mathcal{A}_- & even & - & 1& h\in\Zh+\frac12,\,h\geqslant -\frac12\\
\hline \Lambda_+ & even & + & 0& h\in\Zh,\,h\geqslant 0\\
\hline \Lambda_- & odd & - & 0& h\in\Zh+\frac12,\,h\geqslant \frac{1}{2}\\ \hline
\end{tabular}
\end{center}
\vspace{-0.5cm}\caption{Grassman properties of the string fields and gauge
parameters in the 0 picture.}\label{tab:Grassman}
\end{table}

Due to the presence of the superconformal ghosts
the NS SFT has some
specific properties as compared with the bosonic theory.
In particular, it is possible to restrict the
string fields to be in a smaller space \cite{review}.

\subsection{Restriction of  String Fields.}
Let us first  formulate the restriction scheme
for GSO$+$ sector, then we generalize this scheme
to the GSO$-$ sector.
We decompose the string field $\Ac\equiv\Ac_+$
according to the $\phi$-charge $q$:
$$
\Ac=\sum_{q\in\Zh}\Ac_q,\qquad \Ac_q\in V_q,
$$
where
\begin{equation}
[j_0,\,\Ac_q]=q\Ac_q \quad\text{with}\quad j_0=\frac{1}{2\pi i
}\oint d\zeta\,\pd\phi(\zeta).
\end{equation}

The BRST charge $Q_B$ has also  the natural decomposition
over $\phi$-charge (see Appendix~\ref{app:notations}):
\begin{equation}
Q_B=Q_0+Q_1+Q_2.
\end{equation}
Since $Q_B{}^2=0$ we get the identities:
\begin{equation}
Q_0{}^2=0,\quad\{Q_0,\,Q_1\}=0,\quad\{Q_1,\,Q_2\}=0,
\quad Q_2{}^2=0 \quad\text{and}\quad
\{Q_0,\,Q_2\}+Q_1{}^2=0.
\end{equation}
The non-chiral inverse double step picture changing operator
$Y_{-2}$ has $\phi$-charge equal to $-4$.
Therefore to be non zero the expression
in the brackets $\la Y_{-2}|\dots\ra$  must have
$\phi$-charge equal to $+2$. Hence the quadratic $S_2$ and cubic $S_3$
terms of the GSO$+$ part of the action (\ref{action}) read:
\begin{subequations}
\begin{align}
S_2&=-\frac{1}{2}\sum_{q\in\Zh}\la Y_{-2}|\Ac_{2-q},\,Q_0\Ac_q\ra
-\frac{1}{2}\sum_{q\in\Zh}\la Y_{-2}|\Ac_{1-q},\,Q_1\Ac_q\ra
-\frac{1}{2}\sum_{q\in\Zh}\la Y_{-2}|\Ac_{-q},\,Q_2\Ac_q\ra.
\label{resA}
\\
S_3&=-\frac{1}{3}\sum_{q,\,q^{\prime}\in\Zh}
\la Y_{-2}|\Ac_{2-q-q^{\prime}},\Ac_{q^{\prime}},\Ac_q\ra.
\end{align}
\label{resQact}
\end{subequations}
 We see that all the fields $\Ac_q$, $q\ne 0,1$, give linear
contribution only to the quadratic action \eqref{resA}.
We propose to exclude fields that at the given  level produce only linear contribution
to the free action. This is similar to a proposal described by eq.(3.4)
in \cite{PTY,AM}, which was obtained as a consequence of
the existence of the nontrivial kernel of operator $Y_{-2}$. We will
consider the action
\eqref{resQact}  with fields that belong to spaces
$V_0$ and $V_1$ only.
To make this prescription meaningful we have to check that
the restricted action is gauge invariant. (The complete analysis of
this issue will be presented in \cite{review}).

The action restricted to  subspaces $V_0$ and $V_1$
takes the form
\begin{subequations}
\begin{align}
S_{2,\,\text{restricted}}&=-\frac{1}{2}\la Y_{-2}|\Ac_{0},\,Q_2\Ac_0\ra
-\la Y_{-2}|\Ac_{0},\,Q_1\Ac_1\ra
-\frac{1}{2}\la Y_{-2}|\Ac_{1},\,Q_0\Ac_1\ra,
\label{act-2}
\\
S_{3,\,\text{restricted}}&=
-\la Y_{-2}|\Ac_{0},\Ac_{1},\Ac_1\ra.
\end{align}
\label{act}
\end{subequations}
The action \eqref{act} has a nice structure.
One sees that since the charge $Q_2$ does not contain
zero modes of the stress energy tensor
the fields $\Ac_0$ play a role of auxiliary fields.
On the contrary,
all  fields $\Ac_1$ are physical ones, i.e. they
have non-zero kinetic terms.

Let us now check that the action \eqref{act} has
gauge invariance. The GSO$+$ part
of the action (\ref{action}) is gauge invariant under
\begin{equation}
\delta\Ac=Q_B\Lambda+[\Ac,\,\Lambda],
\label{gaugeP}
\end{equation}
where $\Lambda\equiv\Lambda_+$.
But after the restriction to the space $V_0\oplus V_1$
it might be lost.
Decomposing the gauge parameter $\Lambda$ over
$\phi$-charge $\Lambda=\sum\limits_q\Lambda_q$ we rewrite the
gauge transformation \eqref{gaugeP} as:
\begin{equation}
\delta\Ac_q=Q_0\Lambda_q+Q_1\Lambda_{q-1}+Q_2\Lambda_{q-2}
+\sum_{q^{\prime}\in\Zh}[A_{q-q^{\prime}},\,\Lambda_{q^{\prime}}].
\label{gaugeR}
\end{equation}
Assuming that $\Ac_q=0$ for $q\neq0,1$
from \eqref{gaugeR} we get
\begin{subequations}
\begin{align}
\delta\Ac_{-2}&=Q_0\Lambda_{-2};
\\
\delta\Ac_{-1}&=Q_0\Lambda_{-1}+[\Ac_0,\,\Lambda_{-1}]
+Q_1\Lambda_{-2}+[\Ac_1,\,\Lambda_{-2}];
\\
\delta\Ac_{0}&=Q_0\Lambda_{0}+[\Ac_0,\,\Lambda_{0}]
+Q_1\Lambda_{-1}+[\Ac_1,\,\Lambda_{-1}]
+Q_2\Lambda_{-2};
\\
\delta\Ac_{1}&=Q_0\Lambda_{1}+[\Ac_0,\,\Lambda_{1}]
+Q_1\Lambda_{0}+[\Ac_1,\,\Lambda_{0}]
+Q_2\Lambda_{-1};
\\
\delta\Ac_{2}&=Q_1\Lambda_{1}+[\Ac_1,\,\Lambda_{1}]
+Q_2\Lambda_{0};\label{da2}
\\
\delta\Ac_{3}&=Q_2\Lambda_{1}.
\end{align}
\label{restr}
\end{subequations}
To make the restriction consistent with transformations
(\ref{restr}) the variations of the fields
$\Ac_{-2},\,\Ac_{-1},\,\Ac_{2}$ and $\Ac_3$ must be zero. Since
the gauge parameters cannot depend on string fields we must put
$\Lambda_{-2}=\Lambda_{-1}=\Lambda_{1}=0$. So we are left with the
single parameter $\Lambda_0$, but to have zero variation of $\Ac_2$ we
must require in addition $Q_2\Lambda_0=0$. Therefore the gauge
transformations take the form
\begin{equation}
\begin{split}
\label{resgt} \delta\Ac_0&=Q_0\Lambda_0+[\Ac_0,\,\Lambda_0],
\\
\delta\Ac_1&=Q_1\Lambda_0+[\Ac_1,\,\Lambda_0],
\quad\text{with}\quad Q_2\Lambda_0=0.
\end{split}
\end{equation}

It is easy to check that (\ref{resgt}) form a closed algebra. It
is also worth to note that  the restriction $Q_2\Lambda_0=0$
leaves the gauge transformation of the massless vector field
unchanged. The complete investigation of the influence of the
restriction condition on the gauge transformations of physical
fields will be presented in \cite{review}.

 It is straightforward to extend the
 above restriction  to GSO$-$ sector. It can
be checked that one gets the same restriction on GSO$-$ gauge
parameter $\Lambda_{0,-}$:
$$
Q_2\Lambda_{0,-}=0.
$$

Let us note that in \cite{ABKM} we have used implicitly the
restricted action and in the  present paper
all calculations are performed
for this restricted action.

\subsection{Gauge Symmetry on Constant Fields.}
In this section we restrict our attention to scalar fields at zero momentum,
which are relevant for calculations of Lorentz-invariant vacuum.
The zero-momentum scalar string fields $\Ac_+$ and $\Ac_-$
can be expanded as
\begin{equation}
\Ac_+ = \sum_{i = 0}^{\infty}\phi^i\Phi_{i}\qquad\text{and}\qquad
\Ac_-= \sum_{a = 0}^{\infty}t^a\mathrm{T}_{a},
\end{equation}
where conformal operators $\Phi_{i}$ and $\mathrm{T}_{a}$ are taken at zero
momentum
and $\phi^i$ and $t^a$ are  constant  scalar fields.
The action \eqref{action} for the component
fields $\phi^i$, $t^a$ is a  cubic polynomial
of the following form
\begin{equation}
S = -\frac{1}{2}\sum _{i,j} \mathscr{M}_{ij}\phi^i\phi^j
-\frac{1}{2}\sum_{a,b} \mathscr{F}_{ab}t^at^b
-\frac{1}{3}\sum _{i,j,k} \mathscr{G}_{ijk}\phi^i\phi^j\phi^k
+\sum_{i,a,b} \mathscr{G}_{iab}\phi^it^at^b,
\label{action-mat}
\end{equation}
where
\begin{subequations}
\begin{alignat}{2}
\mathscr{M}_{ij} &= \la Y_{-2}| \Phi_i,Q_B\Phi_j\ra,
&\qquad\qquad
\mathscr{G}_{ijk} &= \la Y_{-2}| \Phi_i,\Phi_j, \Phi_k\ra,
\\
\mathscr{F}_{ab} &= \la Y_{-2}| \mathrm{T}_a,Q_B\mathrm{T}_b\ra, &
\mathscr{G}_{iab} &= \la Y_{-2}| \Phi_i,\mathrm{T}_a,
\mathrm{T}_b\ra.
\end{alignat}
\end{subequations}

For the sake of simplicity we consider the gauge transformations
with GSO$-$ parameter $\Lambda_-$ equal to zero.
The scalar constant
gauge parameters $\{\delta\lambda^{\alpha}\}$ are the components of a ghost number zero
GSO$+$ string field
\begin{equation}
\Lambda_{+} =
\sum_{\alpha} \delta\lambda^{\alpha}\Lambda_{+,\alpha} .
\end{equation}
Assuming that the basis $\{\Phi_j,\,\mathrm{T}_b\}$ is complete
we write the following identities:
\begin{subequations}
\label{def-J}
\begin{align}
Q_B\Lambda_{+,\alpha}
=\sum \mathscr{V}^i_{\alpha}\Phi _i,\\
\Phi_j\star\Lambda_{+,\alpha}-\Lambda_{+,\alpha}\star\Phi_j&=
\sum_{i}\mathscr{J}^i{}_{j\alpha}\Phi_i,
\\
\mathrm{T}_b\star\Lambda_{+,\alpha}-\Lambda_{+,\alpha}\star\mathrm{T}_b&=
\sum_{a}\mathscr{J}^a{}_{b\alpha}\mathrm{T}_a.
\end{align}
\end{subequations}
The variations of the component fields $\phi^i$ and $t^a$
with respect to the gauge transformations \eqref{gauge}
generated by  $\delta\lambda^{\alpha}$ can be
expressed in terms of the ``structure constants''
\begin{subequations}
\begin{align}
\delta \phi^i &\equiv
\delta_0\phi^i +
\delta _1\phi^i =
(\mathscr{V}^i_{\alpha} +  \mathscr{J}^i{}_{j\alpha} \phi^j)
\delta\lambda^{\alpha},
\label{ptrans}
\\
\delta t^a &\equiv
\delta_1 t^a =\mathscr{J}^a{}_{b\alpha} t^b\delta\lambda^{\alpha}.
\label{strans}
\end{align}
\label{trans}
\end{subequations}
The constants $\mathscr{V}^i_{\alpha}$ solve the zero vector
equation for the matrix $\mathscr{M}_{ij}$:
\begin{equation}
\mathscr{M}_{ij}\mathscr{V}^j_{\alpha}=0
\label{zero}
\end{equation}
and therefore the quadratic action is always invariant with respect to free
gauge transformations.

In the bosonic case one deals only with the gauge transformations
of the form \eqref{ptrans}
and finds $\mathscr{J}^i{}_{j\alpha}$ \cite{Taylor} using
an explicit form of $\star$-product \cite{Witten}
in terms of the Neumann functions
\cite{GrJe}. In our case it is more suitable to employ the
conformal field theory calculations using
the following identity:
\begin{equation}
\la Y_{-2}| \Phi_1,\,\Phi_2\star\Phi_3\ra=\la Y_{-2}| \Phi_1,\,\Phi_2,\,\Phi_3\ra.
\label{star}
\end{equation}
To this end it is helpful to
use a notion of dual conformal operator.
Conformal operators $\{\tilde{\Phi}^i,\,\tilde{\mathrm{T}}^a\}$ are
called dual to the operators $\{\Phi_j,\,\mathrm{T}_b\}$ if the following
equalities hold
\begin{equation}
\la Y_{-2}| \tilde {\Phi}^i,\,\Phi_j\ra=\delta^i{}_{j}
\qquad\text{and}\qquad
\la Y_{-2}| \tilde {\mathrm{T}}^a,\,\mathrm{T}_b\ra=\delta^a{}_{b}.
\label{Def_d}
\end{equation}

Using \eqref{star}, \eqref{Def_d} and \eqref{def-J} we can express the
structure constants $\mathscr{J}^i{}_{j\alpha}$
and $\mathscr{J}^a{}_{b\alpha}$ in terms of the correlation functions:
\begin{subequations}
\begin{align}
\mathscr{J}^i{}_{j\alpha}&=
\la Y_{-2}| \tilde {\Phi}^i,\,\Phi_j,\,\Lambda_{+,\alpha}\ra
-\la Y_{-2}| \tilde {\Phi}^i,\,\Lambda_{+,\alpha},\,\Phi_j\ra,
\\
\mathscr{J}^a{}_{b\alpha}&=
\la Y_{-2}| \tilde {\mathrm{T}}^a,\,\mathrm{T}_b,\,\Lambda_{+,\alpha}\ra
-\la Y_{-2}| \tilde {\mathrm{T}}^a,\,\Lambda_{+,\alpha},\,\mathrm{T}_b\ra.
\end{align}
\label{str-const}
\end{subequations}
In the next section these formulae are used to write down
gauge transformations explicitly.


%% file: GFC-calc.tex

We have computed \cite{ABKM}
the restricted action \eqref{act-2} up to
level $(2,6)$. The relevant conformal fields
with $\phi$-charge $1$ and $0$ are
\begin{subequations}
\begin{alignat}{3}
\Phi_0\equiv U&=\,c
&\qquad \Phi_3\equiv V_3&=\,cT_{\eta\xi}
&\qquad \Phi_6\equiv V_6&=\,T_F\eta e^{\phi}
\\
\Phi_1\equiv V_1&=\,\partial^2c
& \Phi_4\equiv V_4&=\,cT_{\phi}
& \Phi_7\equiv V_7&=\,bc\pd c
\\
\Phi_2\equiv V_2&=\,cT_B
& \Phi_5\equiv V_5&=\,c\partial^2\phi
& \Phi_8\equiv V_8&=\,\pd c\pd\phi
\\
&&\mathrm{T}_0&=\frac 14\,e^{\phi}\eta
\end{alignat}
\label{operators}
\end{subequations}
with $\phi^i=\{u,v_1,\dots,v_8\}$ and $t^a=\{t\}$.
For this set of fields we have got
\begin{align}
S_2^{(2,4)}&=u^2+\frac{1}{4}t^2+(4v_1-2v_3-8v_4+8v_5+2v_7)u
\nonumber
\\
&+4v_1^2+\frac{15}{2}v_2^2+v_3^2+\frac{77}{2}v_4^2+22v_5^2+10v_6^2
+8v_1v_3-32v_1v_4+24v_1v_5+4v_1v_7
\nonumber
\\
&-16v_3v_4+4v_3v_5-2v_3v_7+12v_3v_8-52v_4v_5-8v_4v_7-20v_4v_8
+8v_5v_7+8v_5v_8
\nonumber
\\
&+(-30v_4+20v_5+30v_2)v_6+4v_7v_8,
\label{S24}
\\
S_3^{(2,6)}&=\left(\frac{1}{3\gamma ^2}u+\frac{9}{8}v_1-\frac{25}{32}v_2-\frac{9}{16}v_3-\frac{59}{32}v_4
+\frac{43}{24}v_5+\frac{2}{3}v_7
\right)t^2
\nonumber
\\
&+\left(-\frac{40\gamma }{3}u-45\gamma ^3v_1
+\frac{45\gamma ^3}{4}v_2+\frac{45\gamma ^3}{2}v_3+\frac{295\gamma ^3}{4}v_4
-\frac{215\gamma ^3}{3}v_5-\frac{80\gamma ^3}{3}v_7\right)v_6^2.
\label{S34}
\end{align}

Here $\gamma=\frac4{3\sqrt{3}}$. There is no gauge transformation at level zero. At level 2 the gauge
parameters are  zero picture conformal fields
with ghost number 0 and the weight $h=1$, see Table 1.
 There are two such conformal
fields with $0$ $\phi$-charge:
$bc$ and $\pd\phi$, i.e.
on the conformal language the gauge parameter $\Lambda_+$
with  the weight $1$ is of the form
\begin{equation}
\Lambda_+=\delta\lambda_1\,bc+\delta\lambda_2\pd\phi.
\label{lh1CFT}
\end{equation}
The zero order gauge transformation (\ref{gaugeP})
of level $2$ fields has the form
\begin{multline}
\delta_0\Ac_+(w)\equiv Q_B\Lambda_+(w)=\\
(-\delta\lambda_2+\frac{3}{2}\delta\lambda_1)\pd^2 c(w)
+\delta\lambda_1\,cT_B(w)
+\delta\lambda_1\,cT_{\xi\eta}(w)
+\delta\lambda_1\,cT_{\phi}(w)
+\delta\lambda_2\,c\pd^2\phi(w)
\\
-\delta\lambda_2\,\eta e^{\phi}T_F(w)
+\delta\lambda_1\,bc\pd c(w)
+\delta\lambda_2\,\pd c\pd\phi(w)
+\frac{1}{4}(\delta\lambda_1-2\delta\lambda_2)\,b\eta\pd\eta e^{2\phi(w)}.
\end{multline}
We see that in accordance with (\ref{da2}) one gets the field $\Phi_9 =
b\eta\pd\eta e^{2\phi}$ from the sector with $q=2$.
Imposing the condition $Q_2 \Lambda_+ =0$
we exclude this field from the consideration, since
\begin{equation*}
Q_2\Lambda_+
=\frac{1}{4}(\delta\lambda_1-2\delta\lambda_2)
b\eta\pd\eta e^{2\phi}(w)=0.
\end{equation*}
This equality yields
\begin{equation}
\delta\lambda_1=2\delta\lambda_2\equiv 2\delta\lambda.
\label{lambda}
\end{equation}
We are left with
the following zero order gauge transformations
of the restricted  action on level $2$:
\begin{alignat}{3}
\delta _0v_1&=2\delta\lambda,
&\qquad\qquad
\delta _0v_4&=2\delta\lambda,
&\qquad\qquad
\delta _0v_7&=2\delta\lambda,
\nonumber
\\
\delta _0v_2&=2\delta\lambda,&
\delta _0v_5&=\delta\lambda,&
\delta _0v_8&=\delta\lambda,
\label{rd2}
\\
\delta _0v_3&=2\delta\lambda,&
\delta _0v_6&=-\delta\lambda,&
\delta _0u&=0.
\nonumber
\end{alignat}
Transformations \eqref{rd2} give
the vector
$\mathscr{V}^{i}_1\equiv \mathscr{V}^{i}$ in \eqref{ptrans}
in the form
\begin{equation}
\mathscr{V}^{i}=\{0,2,2,2,2,1,-1,2,1\}.
\label{V9}
\end{equation}

One can check that the quadratic action
at level $(2,4)$ (\ref{S24}) is invariant
with respect to this transformation
\begin{equation}
\delta _0 S_2=\delta\lambda \sum _{i=1}^{9}
\frac{\partial S_2}{\partial \phi^{i}}\mathscr{V}^i=0,
\end{equation}
or in other words
$9$-component vector $\mathscr{V}^i$ \eqref{V9}
is the zero vector of the matrix $\mathscr{M}_{ij}$ defined
by \eqref{S24}.

Now we would like to find the nonlinear terms in the transformations
(\ref{trans}). At  level $2$ we have
$\mathscr{J}^i_{j1}=\mathscr{J}^i_{j}$.The dual operators
\eqref{Def_d} to the operators \eqref{operators} are the following
\begin{subequations}
\begin{alignat}{2}
\tilde{\Phi}^1&=\frac{1}{16}\,\eta\partial\eta\bigl[ 1
+\partial b c\bigr]e^{2\phi},
&\qquad
\tilde{\Phi}^5&=-\frac{1}{16}\,\eta\partial\eta\bigl[4-\partial^2\phi
+2\,\partial\phi\partial\phi\bigr] e^{2\phi},
\\
\tilde{\Phi}^2&=\frac{1}{60}\,\eta\partial\eta e^{2\phi}T_B,
&\qquad
\tilde{\Phi}^6&=-\frac{1}{20}\,cT_F\partial\eta e^{\phi},
\\
\tilde{\Phi}^3&=\frac{1}{48}\,\bigl[\partial\eta\partial^2\eta
-6\,\eta\partial\eta \bigr]e^{2\phi},
&\qquad
\tilde{\Phi}^7&=\frac{1}{8}\,\eta\partial\eta\bigl[1
+b\partial c\bigr] e^{2\phi},
\\
\tilde{\Phi}^4&=-\frac{1}{8}\,\eta\partial\eta\partial\phi\partial\phi e^{2\phi},
&\qquad
\tilde{\Phi}^8&=\frac{1}{8}\,\eta\partial\eta bc\partial\phi e^{2\phi},
\end{alignat}
\vspace{-8mm}
\begin{alignat}{2}
\tilde{\Phi}^0&=\frac{1}{8}\,\eta\partial\eta\bigl[
6-\partial (b c)-2\partial^2\phi\bigr] e^{2\phi}
+\frac{1}{48}\,\partial\eta\partial^2\eta e^{2\phi},
&\qquad
\tilde{\mathrm{T}}^0&=\frac{1}{2}\,c e^{\phi}\partial\eta.~~~~
\end{alignat}
\label{dual}
\end{subequations}
It is straightforward to check that
\begin{equation}
\la Y_{-2}|\tilde{\Phi}^{j},\,\Phi_i\ra=\delta^j{}_{i}
\qquad\text{and}\qquad
\la Y_{-2}| \tilde{\mathrm{T}}^{0},\,\mathrm{T}_{0}\ra=1.
\end{equation}

We find the coefficients $\mathscr{J}^i_j$ in (\ref{trans}) up to
level $(2, 4)$ and this gives the following gauge transformations
\begin{subequations}
\begin{align}
\delta_1 u&=[(-\frac{82}{3}\gamma^3+32\gamma)v_1-\frac{16}{3}\gamma^3v_4
+(-19\gamma^3+16\gamma)v_5
+(-\frac{73}{3}\gamma^3+16\gamma)v_7
\nonumber
\\
&~~~~~~~~~~~~~+(\frac{154}{3}\gamma^3-32\gamma)v_8]~\delta\lambda,
\\
\delta_1 t&=\frac{4}{3}t~\delta\lambda,
\\
\delta_1 v_1&=[(-\frac{27}{2}\gamma^3+\frac{8}{3}\gamma)v_1
+(-\frac{11}{4}\gamma^3+\frac{4}{3}\gamma)v_5
+(-\frac{17}{12}\gamma^3+\frac{4}{3}\gamma)v_7
+(\frac{3}{2}\gamma^3-\frac{8}{3}\gamma)v_8]~\delta\lambda,
\\
\delta_1 v_2&=[-\frac{5}{3}\gamma^3v_1-\frac{5}{6}\gamma^3v_5-\frac{5}{6}\gamma^3v_7
+\frac{5}{3}\gamma^3v_8]~\delta\lambda,
\\
\delta_1 v_3&=[(\frac{17}{3}\gamma^3-\frac{16}{3}\gamma)v_1
+(\frac{17}{6}\gamma^3-\frac{8}{3}\gamma)v_5
+(\frac{17}{6}\gamma^3-\frac{8}{3}\gamma)v_7+(-\frac{17}{3}\gamma^3+
\frac{16}{3}\gamma)v_8]~\delta\lambda,
\\
\delta_1 v_4&=[-\frac{5}{3}\gamma^3v_1+\frac{32}{3}\gamma^3v_4-\frac{37}{6}\gamma^3v_5
-\frac{5}{6}\gamma^3v_7+\frac{5}{3}\gamma^3v_8]~\delta\lambda,
\\
\delta_1 v_5&=[-\frac{4}{3}\gamma u+(\frac{61}{6}\gamma^3-\frac{32}{3}\gamma)v_1
+\frac{25}{8}\gamma^3v_2
-\frac{5}{12}\gamma^3v_3+\frac{481}{24}\gamma^3v_4
\nonumber
\\
&~~~~~~~~~~~
+(-\frac{31}{6}\gamma^3-\frac{16}{3}\gamma)v_5
+(\frac{14}{3}\gamma^3-\frac{16}{3}\gamma)v_7+(-\frac{52}{3}\gamma^3+
\frac{32}{3}\gamma)v_8]~\delta\lambda,
\\
\delta_1 v_6&=
-\frac{4}{3}\gamma^3v_6~\delta\lambda,
\\
\delta_1 v_7&=[\frac{16}{3}\gamma u+(\frac{38}{3}\gamma^3+\frac{16}{3}\gamma)v_1
-\frac{25}{2}\gamma^3v_2+\frac{5}{3}\gamma^3v_3
-\frac{193}{6}\gamma^3v_4
\nonumber
\\
&~~~~~~~~~~~~+(\frac{70}{3}\gamma^3+\frac{8}{3}\gamma)v_5
+(8\gamma^3+\frac{8}{3}\gamma)v_7
+(16\gamma^3-\frac{16}{3}\gamma)v_8]~\delta\lambda,
\\
\delta_1 v_8&=[\frac{4}{3}\gamma u+\frac{43}{6}\gamma^3v_1-\frac{25}{8}\gamma^3v_2
+\frac{5}{12}\gamma^3v_3+\frac{95}{24}\gamma^3v_4
+\frac{11}{6}\gamma^3v_5+4\gamma^3v_7]~\delta\lambda,
\end{align}
\end{subequations}
where $\gamma=\frac{4}{3\sqrt{3}}\approx 0.770$.


%% file: GFC-glts.tex
As it has been mentioned above (see \eqref{zero}) the
quadratic restricted
action (\ref{act-2})
 is invariant with respect to the free gauge transformation
\eqref{rd2}.

In contrast to the bosonic case \cite{Taylor}
already the first order gauge
invariance is broken by the level truncation scheme.
Explicit calculation shows that
\begin{equation}
\delta S|_{\text{first order}}\equiv
\delta_1S_2^{(2,4)}+\delta_0S_3^{(2,6)}=-\frac13t^2\delta\lambda+
\{\text{quadratic terms in } v_i\}\delta\lambda.
\end{equation}
Note that the terms in the braces belong to level $6$
and therefore we neglect them.

The origin of this breaking is in the presence of non-diagonal
terms in the quadratic action \eqref{S24}. More precisely, in the bosonic
case the operators with different weights are orthogonal
to each other with respect to the odd bracket,
while in the fermionic string due
to the presence of
operator $Y_{-2}$ this orthogonality is violated.
Indeed, the substitution of $\{\phi^i\}=\{u,v_i,\,w^I\}$
and $\{t^a\}=\{t,\,\tau^A\}$ (here by $w^I$ and $\tau^A$
we denote higher level fields) into the action \eqref{action-mat}
yields
\begin{subequations}
\begin{align}
S_2 &= \frac14t^2 -\sum_A \mathscr{F}_{tA}t\tau^A
-\frac12\sum_{A,B} \mathscr{F}_{AB}\tau^A\tau^B
-\frac12\sum_{i,j} \mathscr{M}_{ij}\phi^i\phi^j,
\\
S_3&=-\frac13\sum _{i,j,k}\mathscr{G}_{ijk}\phi^i\phi^j\phi^k
+\sum_{i} \mathscr{G}_{itt}\phi^it^2
+\sum_{i,A} (\mathscr{G}_{itA}+\mathscr{G}_{iAt})\phi^i t\tau^A
+\sum_{i,A,B} \mathscr{G}_{iAB}\phi^i\tau^A\tau^B.
\end{align}
\end{subequations}
For gauge transformations (\ref{gaugeP}) with $\Lambda_+$
in the form (\ref{lh1CFT}), (\ref{lambda}) we have
\begin{subequations}
\begin{align}
\delta_0 t&=0,\quad\delta_0\tau^A=0,\quad\text{and}\quad \delta_0w^I=0
\label{0tr}
\\
\delta_1t&=\frac43t\delta\lambda
\quad\text{and}\quad
\delta_1\tau^A= (\mathscr{J}^A{}_{t} t+\mathscr{J}^A{}_{B} \tau^B)
\delta\lambda
\label{1tr}
\end{align}
\end{subequations}

In the first order
gauge transformation of the action produces the following quadratic in $t^a$ terms
\begin{multline}
\label{exact}
\left.(\delta_1 S_2+\delta_0S_3)\right|_{t^at^b-\text{terms}}=
\left.\left(\frac{\pd S_2}{\pd t}\delta_1 t
+\frac{\pd S_2}{\pd \tau^A}\delta_1\tau^A
+\frac{\pd S_3}{\pd u}\delta_0 u
+\frac{\pd S_3}{\pd v_i}\delta_0v_i\right)\right|_{t^at^b-\text{terms}}
\end{multline}
Here we take into account that due to (\ref{0tr})
\begin{equation}
\delta_0 S_3|_{\text{contributions from higher levels}}=0
\end{equation}
The exact gauge invariance means that (\ref{exact})
 equals to zero. In the presence of non-diagonal terms in $S_2$ we have
\begin{equation}
\frac{\partial S_2}{\partial \tau^A}\delta_1 \tau^A
=\left[-\mathscr{F}_{tA}\mathscr{J}^A{}_{t}t^2
-(\mathscr{F}_{tA}\mathscr{J}^A{}_{B}
+\mathscr{F}_{AB}\mathscr{J}^A{}_t)t\tau^B-
\mathscr{F}_{AB}\mathscr{J}^B{}_{C}\tau^A\tau^C\right]\delta\lambda
\label{next}
\end{equation}
Generally speaking, $\mathscr{F}_{At}\neq 0$
and  (\ref{next}) contains $t^2$ term.
Therefore if we exclude fields $\tau ^A$  from $S_2$
we cannot compensate $\mathscr{G}_{itt}t^2\delta_0\phi^i$ that
breaks the first order gauge invariance.

We can estimate
the contribution of higher level fields $\{\tau^A\}$
to \eqref{next}. Let us consider only $t^2$ terms in
\eqref{exact}:
\begin{equation}
\left(\frac23 t^2-\mathscr{F}_{tA}\mathscr{J}^A{}_t t^2\right)\delta\lambda
+\mathscr{G}_{itt}\mathscr{V}_i\delta\lambda=0.
\label{EQSD}
\end{equation}
One can check that $\mathscr{G}_{itt}\mathscr{V}_i=-1$ and hence
$\mathscr{F}_{tA}\mathscr{J}^A{}_t=-\frac13$.
Therefore we see that the contribution of the higher levels
into equality \eqref{EQSD} is only $33\%$. This gives us a hope that the
gauge invariance rapidly restores as level grows. For bosonic case
this restoration was advocated in \cite{Taylor}.


%% file: GFC-orbits.tex

In \cite{ABKM} we have used a special gauge
\begin{equation}
G(\phi^i)\equiv 3v_2-3v_4+2v_5=0
\label{gaugecond}
\end{equation}

Calculations of the potential
in this gauge are more simple. However, one has to study
the range of  validity of this gauge choice. For example,
in the case of the
bosonic string the Feynman-Siegel gauge is not universal
within the level truncation method
\cite{Taylor}.
To study the validity of the gauge \eqref{gaugecond}
we investigate
the orbits of the gauge group in the level
truncation scheme.
To this end we have to solve the equations
\begin{subequations}
\begin{align}
\frac{d\phi^i(\lambda)}{d\lambda}&=
\mathscr{V}^i+ \mathscr{J}^i{}_{j}\phi^j(\lambda),
\quad\text{with}\quad\phi^i(0)=\phi^i_0,
\label{eqphi}
\\
\frac{dt^a(\lambda)}{d\lambda}&=
\mathscr{J}^a{}_{b}t^b(\lambda),
\quad\text{with}\quad t^a(0)=t^a_0.
\label{eqt}
\end{align}
\label{equations}
\end{subequations}
To write down explicit solutions of \eqref{eqphi}\eqref{eqt}
we use another basis for $\phi_i$ in which the matrices
$\mathscr{J}$ have the canonical Jordan form.

\subsection{Orbits in Bosonic String Field Theory}
As a simple example of the gauge fixing in the level truncation scheme
let us consider the Feynman-Siegel gauge
at  level $(2,6)$ in bosonic open string field theory.
Here we use notations of \cite{zwiebach}.
Up to level $2$ the string field has the expansion:
\begin{equation}
\Phi =\sum _1^4 \phi^i\Phi_i\qquad\text{with}\quad
\phi=\{t,\,v,\,u,\,w\}
\label{bas_f}
\end{equation}
and
\begin{equation}
\Phi_1=c,~\Phi_2=cT_B,~\Phi_3=\frac12\partial^2c,~\Phi_4=bc\partial c
\label{oper-b}
\end{equation}
The Feynman-Siegel gauge $b_0\Phi=0$ on level $2$ looks like
\begin{equation}
G_{FS}(\phi^i)\equiv \phi^4=0.
\label{FZgauge}
\end{equation}
The dual operators to \eqref{oper-b} are the following
\begin{equation}
\tilde{\Phi}^1=c\partial c,~\tilde{\Phi}^2
=\frac{1}{3}c\partial cT_B,
~\tilde{\Phi}^3=\frac12\partial c\partial^2c,
~\tilde{\Phi}^4=\frac16\partial^3cc.
\end{equation}
Indeed, one can check that
\begin{equation}
\la\tilde{\Phi}^i,\Phi_j\ra=\delta^i_{j}, \quad i,j=1,\dots,4.
\end{equation}
The gauge parameter at this level  is:
$$
\Lambda=\delta\lambda_1\Lambda_1,~~~\Lambda_1=bc ,~~~\delta\lambda_1\equiv
\delta\lambda$$
The vector $\mathscr{V}^i$ in \eqref{eqphi} is of the form:
\begin{equation}
\mathscr{V}^i=
\begin{bmatrix}
0\\
1/2\\
-3\\
-1
\end{bmatrix}
\end{equation}
The structure constants of the gauge transformation are given by the
matrix $\mathscr{J}^i{}_{j1}\equiv \mathscr{J}^i{}_{j}$
\begin{equation}
\mathscr{J}^i{}_{j}
=\la\tilde{\Phi}^i,\Phi_j,\Lambda_1\ra
-\la\tilde{\Phi}^i,\Lambda_1,\Phi_j\ra.
\end{equation}
The matrix $\mathscr{J}^i{}_j$ has the following entries
\begin{equation}
[\mathscr{J}^i{}_{j}]=
\begin{bmatrix}
-\frac{1}{\gamma}&\frac{65}{16}\gamma
&\frac{29}{16}\gamma&-\frac{3}{2}\gamma\\
\\
\frac{5}{16}\gamma&-\frac{581}{256}\gamma^3&-\frac{145}{256}\gamma^3
&\frac{15}{32}\gamma^3\\
\\
\frac{11}{16}\gamma&-\frac{715}{256}\gamma^3&-\frac{703}{256}\gamma^3
&-\frac{47}{32}\gamma^3\\
\\
\frac{21}{8}\gamma&-\frac{1356}{128}\gamma^3&\frac{31}{128}\gamma^3&
\frac{63}{16}\gamma^3
\end{bmatrix},
\label{J-bos}
\end{equation}
where $
\gamma=\frac4{3\sqrt{3}}$.
This result coincides with the one obtained in
eq. (9) of \cite{Taylor} with obvious redefinition
of the fields $\phi^i$.

The characteristic polynomial $\mathscr{P}$ of the matrix
$\mathscr{J}^i{}_j$ is
$$
\mathscr{P}(\mathscr{J},\omega)=
\omega^4+\frac{335}{324}\sqrt{3}\,\omega^3
-\frac{3584}{6561}\,\omega^2 +\frac{11869696}{4782969}\sqrt{3}\,\omega
+\frac{819200}{531441}.
$$
The roots
of the characteristic polynomial are
\begin{align}
\{\omega\}=\{-2.565,\,-0.332,\,0.553\pm i1.226\}.
\end{align}
The corresponding four eigenvectors are
\begin{equation}
\nu_{\omega}=
\begin{pmatrix}
0.131\omega^3+1.661\omega^2-0.804\omega+4.655
\\
0.464\omega^3+0.687\omega^2-0.532\omega+2.276
\\
-0.249\omega^3-0.127\omega^2+0.298\omega-1.189
\\
1
\end{pmatrix}.
\label{eigenstate-b}
\end{equation}

\begin{figure}[t]
\centering
\includegraphics[width=380pt]{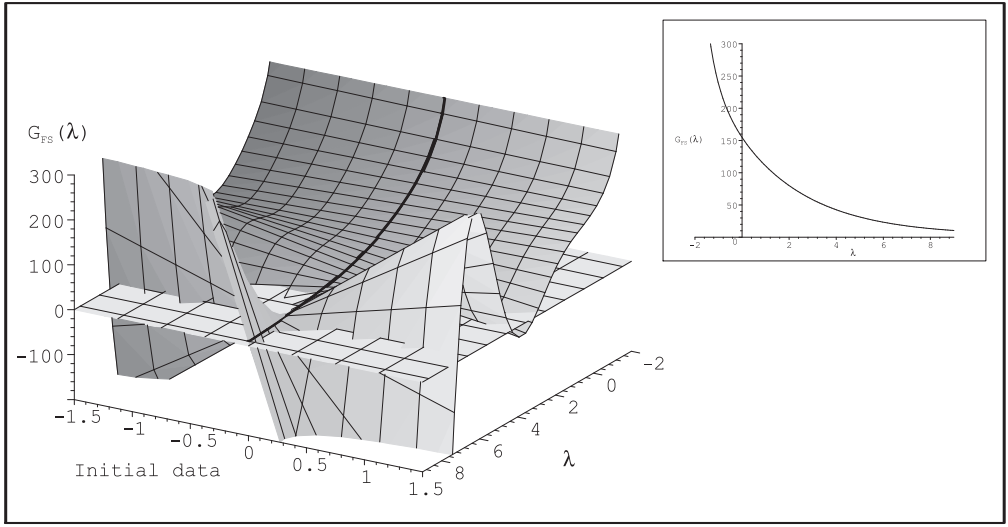}
\caption{Gauge orbits in boson string field theory.}
\label{pic:boson}
\end{figure}

We solve  system \eqref{eqphi} in the basis of these eigenvectors and get
the following dependence of $G_{FS}$ \eqref{FZgauge}
on $\lambda$ (see Figure~\ref{pic:boson}):
\begin{equation}
G_{FS}(\lambda)=
[a\sin{(1.23\lambda)}+b\cos{(1.23\lambda)}]e^{0.553\lambda}
+c e^{-0.332\lambda}
+d e^{-2.57\lambda}+2.89,
\label{FSorbit}
\end{equation}
where
\begin{subequations}
\begin{align}
a&=1.13\,t_0-2.13\,v_0+0.997\,u_0+0.921\,w_0-0.0861,
\\
b&=0.172\,t_0-0.939\,v_0-0.346\,u_0+1.04\,w_0-1.49,
\\
c&=0.0466\,t_0+0.300\,v_0-0.0149\,u_0-0.00958\,w_0-0.741,
\\
d&=-0.218\,t_0+0.639\,v_0+0.360\,u_0-0.0337\,w_0-0.657.
\end{align}
\end{subequations}
If one takes an initial  point $\{t_0,\,v_0,\,u_0,\,w_0\}$
 such  that $a=b=0$ and $c,d\geqslant 0$
then the corresponding gauge orbit never intersects
the surface $G_{FS}(\phi^i)=0$
This situation is depicted in Figure~\ref{pic:boson} by the thick line
on 3-dimensional plot and by 2-dimensional plot.

\subsection{Orbits in Superstring Field Theory}
Performing similar calculations in cubic SSFT we get
the following results.
The characteristic polynomial $\mathscr{P}$ of the
matrix $\mathscr{J}^i{}_j$
is
\begin{equation}
\mathscr{P}(\mathscr{J},\omega)=
\bigl(\omega^4-\frac{1187840}{59049}\omega^2+\frac{451911090176}{3486784401}\bigr)
\bigl(\omega+\frac{256}{729}\sqrt{3}\bigr)
\omega^4
\end{equation}
The eigenvalues of the matrix $\mathscr{J}^j{}_i$
are
\begin{align}
\{\omega\}=\{
0,\, 0,\, 0,\, 0,\, -0.608,\,
\eta
\}\qquad\text{where}\qquad \eta=\pm3.274\pm i0.814
\label{eigenvalues-s}
\end{align}
with the corresponding eigenvectors
\begin{align}
\nu_{0}^{(1)}=
\begin{bmatrix}
1\\ 0\\ 0\\ -5.4\\ 0\\ 0\\ 0\\ 0\\ 0
\end{bmatrix},
\qquad
\nu_{0}^{(2)}=
\begin{bmatrix}
0\\ 0\\ 0\\ 39.3\\ 1\\ 2\\ 0\\ -6\\ -2
\end{bmatrix},
\qquad
\nu_{0}^{(3)}=
\begin{bmatrix}
0\\ 0\\ 1\\ 7.5\\ 0\\ 0\\ 0\\ 0\\ 0
\end{bmatrix},
\qquad
\nu_{0}^{(4)}=
\begin{bmatrix}
0\\ 1\\ 0\\ 155.6\\ 0\\ 0\\ 0\\ -18\\ -8
\end{bmatrix},
\\
\nonumber
\\
\nu_{-0.608}=
\begin{bmatrix}
0\\ 0\\ 0\\ 0\\ 0\\ 0\\ 1\\ 0\\ 0
\end{bmatrix},\qquad
\nu_{\eta}=
\begin{bmatrix}
0.011\eta^3-0.17\eta^2+0.12\eta-1.77
\\
-0.015\eta^3+0.022\eta^2-0.16\eta-0.28
\\
0.5
\\
1
\\
0.038\eta^3+0.26\eta^2+0.39\eta-0.044
\\
-0.03\eta^3+0.044\eta^2+\eta+1.95
\\
0
\\
0.29\eta^3+0.995\eta^2-2.27\eta-7.33
\\
0.12\eta^3+0.54\eta^2-0.13\eta-2.97
\end{bmatrix}.
\label{eigenstate-s}
\end{align}

The solution of \eqref{equations}
yields the gauge orbits of gauge fixing
function \eqref{gaugecond}
\begin{multline}
G(\phi)=[a\sin(0.814\lambda)+b\cos(0.814\lambda)]e^{3.27\lambda}
\\
+[c\sin(0.814\lambda)+d\cos(0.814\lambda)]e^{-3.27\lambda}+
4.15\lambda +f,
\end{multline}
where
\begin{figure}[t]
\centering
\includegraphics[width=380pt]{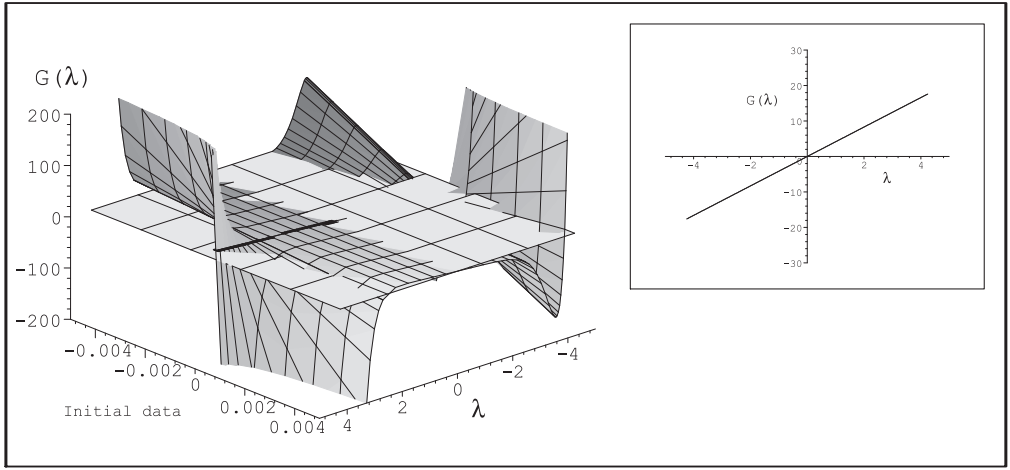}
\caption{Gauge orbits in NS string field theory.}
\label{pic:super}
\end{figure}
\begin{subequations}
\begin{align}
a&=-1.05 u_0 -2.81 v_{1,0} +1.46 v_{2,0} -0.195 v_{3,0}\notag\\
&~~~~~~~~~~+5.82 v_{4,0} -4.81 v_{5,0} -2.13 v_{7,0} +0.614 v_{8,0} -0.177\\
b&=-0.406 u_0 -1.04 v_{1,0} +0.559 v_{2,0} -0.0754 v_{3,0}\notag\\
&~~~~~~~~~~-0.03 v_{4,0} -0.59 v_{5,0} -0.64 v_{7,0} -0.155 v_{8,0} -0.924\\
c&=-0.0933 u_0 -1.38 v_{1,0} +0.130 v_{2,0} -0.0174 v_{3,0}\notag\\
&~~~~~~~~~~+1.06 v_{4,0} -0.572 v_{5,0} + 0.166 v_{7,0} -0.885 v_{8,0} +0.516\\
d&=-0.0636 u_0+ 0.104 v_{1,0} +0.088 v_{2,0} -0.0118 v_{3,0}\notag\\
&~~~~~~~~~~+0.552 v_{4,0} -0.195 v_{5,0} +0.085 v_{7,0} -0.407 v_{8,0} -0.186\\
f&=0.465 u_0+ 0.939 v_{1,0} +2.35 v_{2,0} +0.0867 v_{3,0}\notag\\
&~~~~~~~~~~-3.51 v_{4,0} +2.77 v_{5,0} +0.56 v_{7,0} +0.563 v_{8,0} +1.12
\end{align}
\end{subequations}
and $u_0, v_{i,0}$ are initial data for the corresponding differential
equations \eqref{equations}.

A simple analysis shows that there are no restrictions
on the range of validity of the gauge (\ref{gaugecond})
(see Figure~\ref{pic:super}). The 2-dimensional plot on the Figure~\ref{pic:super}
corresponds to the special initial data $a=b=c=d=0$.

It is interesting to note that there is another gauge which strongly
simplifies the effective potential. Namely, this is the gauge $v_6=0$.
The orbits of this gauge condition have the form
$$
v_6(\lambda)=(1.64+v_{6,0})e^{-0.608\lambda}-1.64
$$
It is evident that this gauge condition is not always reachable and
cannot be used in the calculation of the tachyon potential.


%% file: GFCa-notations.tex
Here we collect
notations that we use in our calculations (for more details see
\cite{fms}).
\renewcommand{\arraystretch}{1.5}
\begin{longtable}[h]{||LL||}
\hline
X_L^{\mu}(z)X_L^{\nu}(w)\sim-\frac{\alpha^{\prime}}{2}\eta^{\mu\nu}
\log(z-w)
&
\psi^{\mu}(z)\psi^{\nu}(w)\sim-\frac{\alpha^{\prime}}{2}\eta^{\mu\nu}
\frac{1}{z-w}
\\
c(z)b(w)\sim b(z)c(w)\sim\frac{1}{z-w}
&
\gamma(z)\beta(w)\sim -\beta(z)\gamma(w)\sim\frac{1}{z-w}
\\
\gamma=\eta e^{\phi}
& \beta=e^{-\phi}\pd\xi
\\
\phi(z)\phi(w)\sim-\log(z-w)
& \xi(z)\eta(w)\sim \eta(z)\xi(w)\sim\frac{1}{z-w}
\\
\hline
\hline
T_B=-\frac{1}{\alpha^{\,\prime}}\pd X\cdot\pd X-\frac{1}{\alpha^{\,\prime}}\pd \psi\cdot\psi
& T_F=-\frac{1}{\alpha^{\,\prime}}\pd X\cdot\psi
\\
T_{bc}=-2b\pd c-\pd bc
& T_{\beta\gamma}=-\frac32\beta\pd\gamma-\frac12\pd\beta\gamma
\\
T_{\phi}=-\frac12\pd\phi\pd\phi-\pd^2\phi
& T_{\eta\xi}=\pd\xi\eta
\\
\hline
\hline
\multicolumn{2}{||L||} {Q_B=Q_0+Q_1+Q_2 }
\\
\multicolumn{2}{||L||}
{Q_0=\frac{1}{2\pi i}\oint d\zeta\,\left[
c(T_B+T_{\phi}+T_{\eta\xi})+bc\pd c\right]}
\\
\multicolumn{2}{||L||}
{Q_1=\frac{1}{2\pi i}\oint d\zeta\,\left[
\frac{1}{\alpha^{\,\prime}}\eta e^{\phi}\psi\cdot\pd X\right]}
\\
\multicolumn{2}{||L||}
{Q_2=\frac{1}{2\pi i}\oint d\zeta\,\left[
\frac14 b\pd\eta\eta e^{2\phi}\right]}
\\
\hline
\hline
Y=4c\pd\xi e^{-2\phi} &\\
\hline
\caption{Notations and OPE-s.}\label{table:2}
\end{longtable}
\renewcommand{\arraystretch}{1}


%% file: GFCa-cmaps.tex
Taylor series of maps \eqref{maps} in the origin:
\begin{subequations}
\begin{align}
F_1(w)&=1+2\gamma w+3\gamma^2 w^2 +\frac{31}{8}\gamma^3
w^3+\frac{39}{8}\gamma^4 w^4 +\frac{813}{128}\gamma^5 w^5+\dots
\\
F_2(w)&=\frac{1}{2}\gamma w-\frac{5}{32}\gamma^3 w^3
+\frac{57}{512}\gamma^5 w^5+\dots
\\
F_{3}(w)&=-1+2\gamma w-3\gamma^2 w^2 +\frac{31}{8}\gamma^3
w^3-\frac{39}{8}\gamma^4 w^4 +\frac{813}{128}\gamma^5 w^5+\dots
\end{align}
\label{mapsa}
\end{subequations}
Here $\gamma=\frac4{3\sqrt{3}}$.

Below we present conformal transformations, necessary
to map dual vertex operators \eqref{dual}
\begin{align*}
(f\circ \eta)(w)&=\df \eta(f)
\\
(f\circ \pd\eta)(w)&=\df^2 \pd \eta(f)+\ddf\eta(f)
\\
(f\circ \pd^2\eta)(w)&=\dddf\eta(f)
+3\df\ddf\pd\eta(f)+\df^3\pd^2\eta(f)
\\
(f\circ T_B)(w)&=\df^2 T_B(f)+\frac{15}{12}\left(
\frac{\dddf}{\df}-\frac{3}{2}\frac{\ddf^2}{\df^2}\right)
\\
(f\circ T_F)(w)&=\df^{3/2} T_F(f)
\\
(f\circ \pd(bc))(w)&=\df^2\pd (bc)(f)
+\ddf b(f)c(f)+\frac{3}{2}\left(\frac{\dddf}{\df}
-\frac{\ddf^2}{\df^2}\right)
\\
(f\circ \pd bc)(w)&=
\df^2\pd b(f)c(f)+2\ddf b(f)c(f)+\frac{5}{6}\frac{\dddf}{\df}
+\frac{1}{4}\frac{\ddf^2}{\df^2}
\\
(f\circ b\pd c)(w)&=
\df^2b(f)\pd c(f)-\ddf b(f)c(f)+\frac{2}{3}\frac{\dddf}{\df}
-\frac{7}{4}\frac{\ddf^2}{\df^2}
\\
(f\circ e^{q\phi})(w)&=\df^{-\tfrac{1}{2}q(q+2)}e^{q\phi(f)}
\\
(f\circ \pd\phi e^{2\phi})(w)&=
\frac{1}{\df^{3}}\pd\phi(f)e^{2\phi(f)}-2\frac{\ddf}{\df^{5}}
e^{2\phi(f)}
\\
(f\circ (\pd\phi)^2 e^{2\phi})(w)&=
\frac{1}{\df^2}\pd\phi(f)\pd\phi(f) e^{2\phi(f)}
-4\frac{\ddf}{\df^4}\pd\phi(f)e^{2\phi(f)}
+\frac{1}{\df^4}\left(-\frac{1}{6}\frac{\dddf}{\df}+
\frac{17}{4}\frac{\ddf^2}{\df^2}\right) e^{2\phi(f)}
\\
(f\circ \pd^2\phi e^{2\phi})(w)&=
\frac{1}{\df^2}\pd^2\phi(f)e^{2\phi(f)}
+\frac{\ddf}{\df^4}\pd\phi(f)e^{2\phi(f)}
+\frac{1}{\df^4}
\left(\frac{3}{2}\frac{\ddf^2}{\df^2}-\frac{5}{3}\frac{\dddf}{\df}
\right) e^{2\phi(f)}
\end{align*}
